\documentclass{paper}
\usepackage{authblk}
\usepackage{hyperref}
\usepackage{graphicx}
\usepackage{fullpage}
\usepackage{enumitem}
\usepackage{lineno}

\makeatletter
\renewcommand\AB@affilsepx{, \protect\Affilfont}
\makeatother

\title{EFT Workshop at Notre Dame}

\author[1,*]{Nick Smith} 
\author[22,*]{Daniel Spitzbart}                       
\author[1,*]{Jennet Dickinson} 
\author[2,*]{Jon Wilson}   
\author[1,*]{Lindsey Gray} 
\author[4,*]{Kelci Mohrman}            
\author[5,*]{Saptaparna Bhattacharya}  
\author[3,*]{Andrea Piccinelli} 
\author[8,*]{Titas Roy}         
\author[9,*]{Garyfallia Paspalaki} 

\author[6,**]{Duarte Fontes}
\author[3,**]{Adam Martin}
\author[7,**]{William Shepherd}

\author[10,**]{Sergio Sánchez Cruz}
\author[11,**]{Dorival Goncalves} 
\author[12,**]{Andrei Gritsan} 
\author[13,**]{Harrison Prosper}

\author[1,**]{Tom Junk}
\author[14,**]{Kyle Cranmer}
\author[15,**]{Michael Peskin}

\author[16,**]{Andrew Gilbert}
\author[17,**]{Jonathon Langford}
\author[18,**]{Frank Petriello}

\author[19,**]{Luca Mantani} 
\author[20,**]{Andrew Wightman}
\author[21,**]{Charlotte Knight}

\author[1]{Prasanth Shyamsundar} 
\author[23]{Aashwin Basnet} 
\author[16]{Giacomo Boldrini} 
\author[3]{Kevin Lannon} 

\affil[1]{Fermilab}
\affil[2]{Baylor U.}
\affil[3]{University of Notre Dame}
\affil[4]{University of Florida}
\affil[5]{Wayne State University}
\affil[6]{Brookhaven National Laboratory}
\affil[7]{Sam Houston State University}
\affil[8]{University of Illinois Chicago}
\affil[9]{Purdue University}
\affil[10]{CERN}
\affil[11]{Oklahoma State University}
\affil[12]{Johns Hopkins University}
\affil[13]{Florida State University}
\affil[14]{University of Wisconsin}
\affil[15]{SLAC}
\affil[16]{Laboratoire Leprince-Ringuet, Ecole Polytechnique, Institut Polytechnique de Paris, Palaiseau, France}
\affil[17]{Imperial College}
\affil[18]{Northwestern University}
\affil[19]{University of Cambridge}
\affil[20]{University of Nebraska Lincoln}
\affil[21]{Imperial College London}
\affil[22]{Boston University}
\affil[23]{Ohio State University}

\affil[*]{editor}
\affil[**]{speaker}

\newcommand{\Dlr}{\mathrel{\raise1.5ex\hbox{$\leftrightarrow$\kern-1em\lower1.5ex\hbox{$D$}}}}
\newcommand{\speaker}[1]{\emph{Speaker:} #1 \par}

\begin{document}

\maketitle

\begin{figure}[ht]
    \centering
    \includegraphics[width=0.7\textwidth]{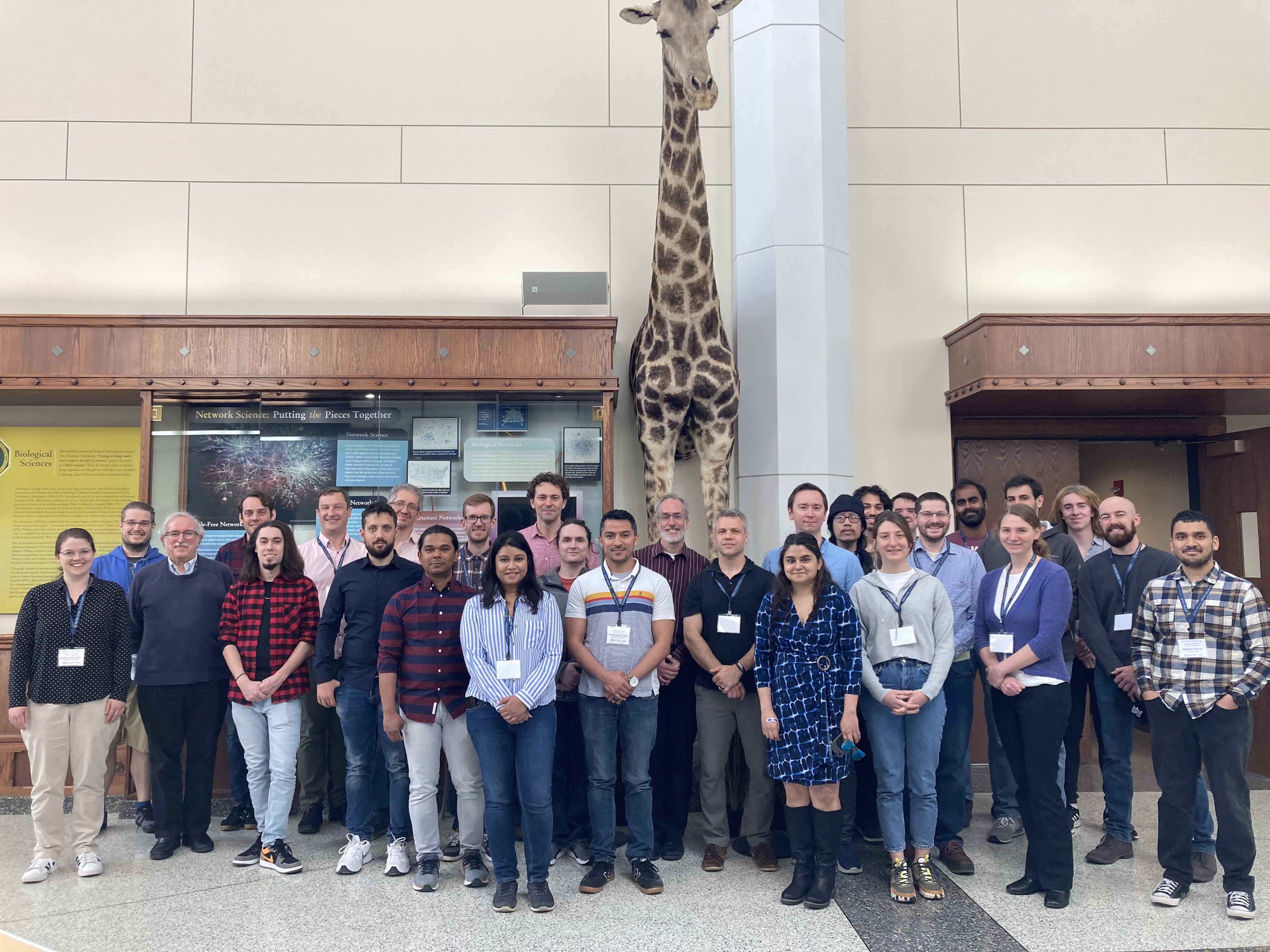}
\end{figure}

\abstract{
The LPC EFT workshop was held April 25-26, 2024 at the University of Notre Dame. The workshop was organized into five thematic sessions: \emph{how far beyond linear} discusses issues of truncation and validity in interpretation of results with an eye towards practicality; \emph{reconstruction-level results} visits the question of how best to design analyses directly targeting inference of EFT parameters; \emph{logistics of combining likelihoods} addresses the challenges of bringing a diverse array of measurements into a cohesive whole; \emph{unfolded results} tackles the question of designing fiducial measurements for later use in EFT interpretations, and the benefits and limitations of unfolding; and \emph{building a sample library} addresses how best to generate simulation samples for use in data analysis. This document serves as a summary of presentations, subsequent discussions, and actionable items identified over the course of the workshop.
}

\clearpage

\section{Introduction}

The LPC\footnote{Large Hadron Collider Physics Center at Fermilab} Effective Field Theory (EFT) workshop was held on April 25-26, 2024 at the University of Notre Dame\footnote{Agenda accessible at \url{https://indico.cern.ch/e/lpceft2024}}. It was the second in a series started at the LPC in September 2023, with a primary goal of growing the community of physicists interested in EFT interpretations of measurements and searches at the CERN Large Hadron Collider (LHC). The workshop included ample time for discussion centered around resolving outstanding practical questions towards performing large-scale EFT combinations within an experiment. The attendees were primarily experimentalists on the CMS experiment and theorists working on EFT-related topics. The desire is for the series to grow in breadth of attendees' backgrounds. Prior to the workshop, an introductory EFT tutorial and hackathon was held. The tutorial introduced EFT concepts and practical tools for new researchers, and the hackathon allowed attendees to work in a self-organized fashion on specific topics, such as generation of a particular sample of events.

This document serves to record what was discussed and what actionable items were identified over the course of the workshop. Speakers were invited to submit a short paragraph summarizing the most critical takeaway messages from their talk to be conveyed to a broader community, and the session conveners were charged with placing the summaries in context by capturing questions and comments raised in subsequent discussions. The sessions were organized into five themes which form the sections of this document: Section~\ref{sec:linear} discusses issues of truncation and validity in interpretation of results with an eye towards practicality; Section~\ref{sec:reconstruction} visits the question of how best to design analyses directly targeting inference of EFT parameters; Section~\ref{sec:logistics} addresses the challenges of bringing a diverse array of measurements into a cohesive whole; Section~\ref{sec:unfolded} tackles the question of designing fiducial measurements for later use in EFT interpretations, and the benefits and limitations of unfolding; and Section~\ref{sec:library} addresses how best to generate simulation samples for use in data analysis.

\section{How far beyond linear?}
\label{sec:linear}

The goal of this session was to explore the considerations  that should be kept in mind when choosing which orders of EFT operators to include in the modeling of EFT predictions for  experimental analyses and to discuss the consequences that this choice has on interpretations. While experimental analyses often include just the linear and dimension-six squared pieces (for practical reasons), this is incomplete. Other contributions (e.g., interference between the SM and dimension-eight contributions, as well as interference between the SM and diagrams with double-insertions of dimension-six vertices) enter at the same $1/\Lambda^4$ order as the dimension-six squared contributions. It is thus important to consider these contributions and work towards more proper methods of handling their effects.

Three main aspects of this topic were explored. Section~\ref{smeft_vs_heft} presents examples of cases where new physics could enter in a way that is not modeled by linear SMEFT contributions. Next, Section~\ref{geosmeft} describes an approach (known as ``geoSMEFT") for identifying and organizing the most relevant higher-order contributions. Section~\ref{dim8_as_uncertainty} discusses how these higher order contributions can be seen as an uncertainty on the linear piece. Finally, Section~\ref{lin_discussion} summarizes the discussion. 

\subsection{SMEFT vs HEFT: When EFT for new physics is not linear}
\label{smeft_vs_heft}
\speaker{Duarte Fontes}
The Higgs Effective Field Theory (HEFT) can be used at the LHC and future colliders to parametrize possible deviations from the Standard Model. Eventual nonzero HEFT coefficients should then be converted into coefficients of specific UV models via a matching procedure. However, this procedure is not unambiguous. In fact, depending on the way in which the different parameters of a certain UV model are assumed to scale, different HEFT expansions (and hence matching relations) are obtained. This happens in such a way that, according to the process and region of the parameter space of the UV model, different matchings should be used to ensure a fast convergence of the HEFT expansion to the results of the UV model. This complicates the interpretation of HEFT measurements in terms of parameters of UV models.

\subsection{Which orders: the geoSMEFT perspective}
\label{geosmeft}
\speaker{Adam Martin}
Using the geoSMEFT reorganization, many processes---including those that are central to the global SMEFT fit program such as the EPWO, $h \to \gamma \gamma$, and diboson processes---are fully calculable to $O(1/\Lambda^4)$. Using these full higher order SMEFT calculations (which, should supersede $|dim6|^2$ approximate calculations whenever they are known), one can study truncation uncertainty. Dimension-8 operators are usually unconstrained by EWPO and can contain different helicity/polarization structure, allowing them to interfere with the SM in scenarios where dimension-6 operators cannot. As a result, they can lead to dramatic differences between the full $O(1/\Lambda^4)$ SMEFT result and the $|dim6|^2$ approximation. The geoSMEFT organization also simplifies the energy vs.~vev\footnote{Higgs field vacuum expectation value} scaling of higher dimensional operators, as all new momentum dependence (novel energy enhanced vertices) are shuffled into process-specific 4+ particle vertices.

\subsection{Truncation and validity: Treating higher order terms as uncertainties}
\label{dim8_as_uncertainty}
\speaker{William Shepherd}

 Exploring specific new physics models and their realizations in EFT frameworks is an interesting exercise, but it doesn't really reflect the state of our current experimental knowledge. A more honest approach to EFT searches must remain model-agnostic, positing only that the SM as we've measured it is the low-energy approximation of some higher-energy theory. This agnosticism suggests that we should not privilege some Wilson Coefficients over others or assume that our framework has been configured such that the interesting effects of this new physics will generically be captured by searches that specifically look for only a small subset of operators, but instead we must be aware that all effects at a given order in the EFT perturbation theory deserve to be treated with the same seriousness.

Taking this perturbation theory seriously necessitates acknowledgement of the presence of higher order calculations than those we can perform, and the treatment of some approximate of them as an error on our understanding of the prediction of the EFT. Without this, we cannot be confident in any constraint we claim on EFT parameters. Luckily, much like QCD offers us the scale dependence as a way to estimate the size of higher-order effects, the EFT also gives us an easy-to-calculate quantity that we can use to get a generic understanding of the size of next-order effects. In an analysis using only dimension-6 operators, this is the square of the amplitude term at order $\frac{1}{\Lambda^2}$. Using this, and scaling to avoid accidental overconfidence (e.g. preventing strong would-be limits at dimension-6 from requiring dimension-8 operators to be small) it is possible to honestly estimate the size of these uncertainties, and yield bounds which are solid constraints on any future model of heavy new physics.

Further details on these ideas can be found in~\cite{Shepherd:2022rsg}. 

\subsection{Discussion}
\label{lin_discussion}

It is evident that higher-order contributions should not be disregarded; however, in many cases, it is not yet practical to fully account for dimension-eight effects. While there is not yet an overarching solution that would be  feasible to apply in all cases,  we summarize the observations and recommendations that resulted from the discussion:
\begin{itemize}
\item The majority of experimental analyses currently include the linear and quadratic dimension-six pieces, without higher order contributions. This approach tends to be utilized because of its practicality. 
\item If there are no other feasible alternatives available, these types of analyses are worthwhile; they provide a method of testing the SM against an alternative hypothesis, so they do have discovery potential. However, analyzers should be aware that the theoretical interpretation of these types of results are limited (not only in the case of a possible sign of new physics, but also for confidence intervals placed on Wilson coefficients in the absence of a signal). 
\item In the absence of a feasible method of properly including all $1/\Lambda^4$ effects, it would be beneficial for analyses to report linear-only results\footnote{It should be noted that there exist cases where a linear term is not present (when the EFT contribution does not interfere with the standard model).} in addition to the quadratic results reported. Many analyses already do this, but it would be beneficial for this to become more standard. 
\item When the full $1/\Lambda^4$ effects cannot be included easily, it may be possible to use the dimension-six quadratic piece to estimate a systematic uncertainty corresponding to the missing $1/\Lambda^4$ terms. While it may be challenging to incorporate this rigorously into a fit in an experimental analysis (as the uncertainty would depend on the Wilson coefficient values, and would need to be handled for each bin) it would be beneficial to invest effort in this area. Results which incorporate such an uncertainty would be more theoretically meaningful than results which aggressively include linear and quadratic dimension-six pieces (without any estimation of uncertainty from the missing $1/\Lambda^4$ terms or the higher-order corrections). Presented together, these conservative and aggressive limits together would help to provide a more useful and interpretable picture of the Wilson coefficient space to be explored. 
\item The geoSMEFT method of organizing dimension-eight contributions provides a promising method to properly handle the most important $1/\Lambda^4$ effects. It would be beneficial for experimental analyses to explore this method, and to use it to begin incorporating proper $1/\Lambda^4$ effects as soon as possible. 
\end{itemize}

The points summarized here do not imply that aggressive analyses (which include linear and quadratic dimension-six contributions) should be abandoned; the discovery potential of these approaches is acknowledged, though the limited interpretability of these approaches is also recognized. The point to be emphasized is that the community should not become complacent in this limited approach. As improved ideas and techniques (such as geoSMEFT) become available, the community should strive to incorporate these into experimental analyses. Regardless of whether future EFT analyses will discover hints of new physics or if limits remain consistent with the standard model prediction, improvements in EFT modeling will strongly benefit the field by improving the interpretability and meaningfulness of LHC EFT results. 

\section{Reconstruction-level results}
\label{sec:reconstruction}
This session focused on the challenges and opportunities posed by performing a search for EFT effects at the reconstruction level (equivalently, detector level). Ideally there would be no loss of information going from truth level to detector level measurements, but to account for this, acceptance and efficiency are taken into account in such measurements but the implementation can be challenging. From designing and picking the most suitable observable for our analyses, to thinking about ways to better utilize and understand existing measurements and statistical methods to derive the results, there is a lot to consider in reconstruction level EFT measurements. To guide our discussion we will be focusing on the following 
\begin{itemize}
    \item Re-interpretability of detector-level measurements Section~\ref{re-interpretability}
    \item Entanglement and Bell’s inequalities with boosted top quarks Section~\ref{entanglement}
    \item Observables for EFT measurements Section~\ref{observables}
    \item Simulation Based Inference Section~\ref{sec:SBI}
\end{itemize}

\subsection{Re-interpretability of detector-level measurements}
\label{re-interpretability}
\speaker{Sergio Sánchez Cruz}
While measurements in dedicated regions of interest with reconstruction level observables have the advantage of reflecting detector effects and portraying a more accurate picture, reinterpreting the results can be challenging. Once a measurement is complete we often lose access to the full statistical model used in the analysis, and any statistical combination becomes impossible. Any changes (like testing the effect of a different EFT operator) or additions to the measurement would also require rerunning the whole analyses, which is not feasible in most cases. The problems posed can be solved if as a collaboration we make it a habit to release the full statistical model used in the analysis, which will make it easier to rerun the analyses or make statistical combinations. Using certain statistical techniques to conduct a post-generation reweighting of simulated samples (Section~\ref{sec:knight}) would also help to study any new operator of interest.

\subsection{Entanglement and Bell’s inequalities with boosted top quarks}
\label{entanglement}
\speaker{Dorival Goncalves}
The LHC offers a unique opportunity to explore quantum correlations, such as entanglement and Bell inequality violation, at the highest energy scale available today. We discuss these quantum correlations using top quark pair production as a model for a two-qubit system, specifically focusing on the semi-leptonic top pair channel, which provides a sixfold increase in statistics and easier reconstruction compared to the dileptonic channel, which was previously used by ATLAS and CMS. While measuring the spin polarization of the hadronic top quark presents challenges, our study demonstrates the feasibility of reconstructing the spin density matrix of the two-qubit system using an optimized hadronic polarimeter. This involves employing jet substructure techniques and reconstruction methods inspired by neural networks to enhance the mapping between subjets and quarks. Our analysis reveals that entanglement can already be observed at a significance level exceeding $5\sigma$ with existing data using this channel. Moreover, the violation of Bell inequalities may be probed at a significance level surpassing $4\sigma$ at the HL-LHC with 3 ab$^{-1}$ of data. Hence, the analysis of this channel can represent the next crucial step for ATLAS and CMS in their exploration of quantum correlations at the LHC.

\subsection{Observables for EFT Measurements}
\label{observables}
\speaker{Andrei Gritsan}
A crucial aspect when devising a detector-level EFT analysis is the proper definition of a set of quantities capable of capturing the peculiarities of the EFT contributions under investigation. The types of suitable observables range from the usual quantities calculated for the SM measurements to the EFT-sensitive, as well as to the optimized observables utilizing advanced matrix-element and machine-learning techniques. 

When building EFT-sensitive observables, it is worth considering that higher-dimension operators typically lead to enhancement at the higher values of four-momenta squared distributions of the particles appearing in the propagators. Therefore, observables based on such calculations or correlated with those quantities become sensitive probes of deviations from the SM. An example of such an observable could be the transverse momentum of reconstructed objects. At the same time, such generic probes may not be sensitive to distinguish multiple operators that all lead to the same enhancement at higher momenta. One example of such a situation is the study of CP-even and CP-odd operators, which may require special CP-sensitive observables to differentiate them. Moving to more complex variables, while the matrix-element calculations guarantee optimal performance from first principles, there are practical limitations to their applications. The most critical limitations are the transfer functions, which are difficult and time-consuming to model. Parton shower and detector effects may confuse and distort the input to matrix elements to such a degree that calculations become impractical. Machine Learning (ML) techniques may come to the rescue in such a case.

Training of machine-learning algorithms is still based on MC samples utilizing the same matrix elements that would be used for optimal discriminants. However, these MC samples reflect the parton shower and detector effects and therefore allow the construction of optimized observables that incorporate these effects. When it comes to performing the training of such an ML model, there are two important aspects to take into account, which are selecting the observables to inject into the learning process and which samples should be used. The matrix-element approach provides answers to both questions, together with an insight into the process of constructing the optimized observables with ML. The first thing to consider is that the input observables should provide full information, which could be simply the four vectors of all particles involved, like in the matrix elements, or better-derived physics quantities that are equivalent to those. In addition, the optimal observables, like in the case of matrix-element-based calculations, can be separated into two categories, which are the set of quantities sensitive to quadratic terms in the amplitude, and the set of variables sensitive to interference terms. The first corresponds to the classic problem of differentiating between two models, and an ML algorithm is trained on two samples corresponding to two alternative models. Training an ML equivalent of the second type of matrix-element-based optimal observables is less obvious, as it requires isolating the interference component. A discriminant trained to differentiate the two models with maximal quantum-mechanical mixing is the best candidate for that goal, where the quadratic term may or may not be removed. Using the two observables jointly would guarantee optimal performance for any size of the contributing operator. 

\subsection{Simulation-based inference}
\label{sec:SBI}
\speaker{Harrison Prosper}
Simulation-based inference (SBI) is in a mature state and high-quality tools exist to implement SBI, notably, the toolkit Madminer~\ref{sec:cranmer}. While it is always helpful to implement SBI oneself, it is worth checking if an approach you wish to use is not already available. Given the mature state of SBI, there is an actual chance to move away from the traditional ML benchmark approach to testing new physics models. It is now computationally feasible to simulate events at a large number of parameter points and either directly construct inference summaries such as point estimates and confidence sets and intervals from simulations or approximate the statistical model $p(X | \theta)$ directly, where $X$ are potential observations and $\theta$ the parameter space of the
theoretical model of reference. It is important to assess the accuracy of the approximations, but if $p(X | \theta)$ can be accurately modeled this has the virtue that all of the standard machinery of frequentist and Bayesian statistical inference can be deployed on a neural network model of $p(X | \theta)$. 

\subsection{Discussion}
When considering observables for EFT measurements, it's crucial to prioritize selecting those that offer the most meaningful insights into the underlying physics. One recommendation is to incorporate a diverse range of observables that capture different aspects of the phenomenon under investigation. This ensures a comprehensive understanding of the system and increases the likelihood of detecting subtle effects indicative of new physics.
In terms of tool selection, it's advisable to leverage versatile tools like MELA, which have evolved to accommodate various generators such as JHUGen, MCFM, and now Madgraph. This adaptability allows for more robust analyses and enhances the compatibility of results across different platforms.
When confronted with a multitude of observables, it's essential to exercise discretion in their selection. Instead of overwhelming analyses with a plethora of variables, focus on identifying a few key observables that are most sensitive to the parameters of interest. This targeted approach not only streamlines the analysis process but also facilitates a clearer interpretation of results.
Furthermore, when constructing likelihood ratios, it's prudent to consider the Neyman-Pearson lemma carefully. While this lemma provides a framework for hypothesis testing, its applicability in multidimensional analyses may vary. It's advisable to assess the suitability of the lemma for the specific context of the analysis and to explore alternative approaches if necessary.

Navigating simulation-based inference techniques involves discerning when to employ matrix-element-driven optimal observables, machine learning, or neither, depending on the analytical context.
\begin{itemize}
\item Prioritize analytical approximation of likelihood whenever possible. This approach fosters a sturdy statistical model and promotes the sharing of likelihoods among researchers. In cases where modeling the likelihood proves impracticable, consider the merits of unbinned analysis techniques.
\item Tackle nuisance parameters by judiciously assigning believable priors and relaxing specific value requirements to facilitate averaging over the priors. However, it's vital to validate the efficacy of this strategy, particularly in higher-dimensional analyses where coverage may vary.
\item
Endeavor to release a CDF of the test statistic distribution, empowering further analysis and interpretation. However, determining the optimal approach between directly approximating the unbinned statistical model or compressing data into a test statistic remains an area requiring additional exploration.
\item Although feasible, combining unbinned and binned analyses necessitates accurate consideration of dependencies between the two statistical models. This requires using compatible parameter spaces to ensure reliable outcomes.
\item Recognize the considerable computational demands associated with simulation-based inference techniques, particularly when producing robust confidence sets. Strategically allocate resources and gradually populate parameter spaces near boundaries to minimize computational expenditures.
\item Focus on identifying independent directions when exploring  EFT parameter spaces. This targeted approach enables effective examination and learning along specific dimensions, enhancing analytical efficiency.
\end{itemize}

In the discussion about the re-interpretability of detector-level measurements, a pertinent question arises concerning the post-analysis re-visitation of full simulations. This practice, once completed, presents a formidable task, given the intricacies involved in re-calibration and re-evaluation. Such challenges underscore not only technical hurdles but also a broader ``sociological problem," wherein entrenched methodologies hinder the exploration of alternative approaches.
However, within this landscape of challenges, avenues for progress emerge. Consider the proposition of presenting a comprehensive likelihood for theoretical scrutiny. While the magnitude of such an undertaking may appear daunting, precedent exists in the form of the published likelihoods in the context of Higgs studies, serving as a testament to the feasibility and benefits of such initiatives. Armed with this foundational framework, theorists gain access to a wealth of insights, fostering deeper exploration and collaborative endeavors.
Amidst these discussions, questions arise regarding disparities observed between helicity-ignorant and helicity-aware reweighting techniques to implement a broad set of EFT scenarios within the same simulated samples. The integration of degrees of freedom and the preservation of spin correlation emerge as pivotal considerations in that respect. Delving deeper, nuances about the consideration of operators and the exploration of phase space come to the fore. A cautious approach is recommended, as drawing conclusive inferences necessitates a thorough understanding of underlying processes and methodological intricacies.

In the discourse surrounding top entanglement and Bell inequalities, several key themes emerge, each offering insights and avenues for further exploration.
Firstly, the observation of improved agreement between CMS measurements of top-entanglement and theoretical predictions underscores the importance of refining modeling techniques, particularly in capturing spin correlations with precision. This highlights the need for ongoing efforts to enhance modeling methodologies to better align experimental observations with theoretical expectations.
Concerns regarding systematic uncertainties prompt reflection on the choice of models for interpreting data, particularly in cases of disagreement between observed data and theoretical predictions. The imperative to ensure robustness and reliability in model selection underscores the critical role of methodological rigor in advancing our understanding of particle interactions.
Considerations surrounding loopholes in Bell inequalities measurements shed light on the inherent challenges in experimental design and interpretation, though collider probes of entanglement can not be used for loophole-free tests of Bell's inequality.
Explorations into the interplay between EFT contributions and top entanglement properties offer fertile ground for future research. Efforts to better exploit EFT frameworks in analyzing angular distributions and spin matrices hold promise for uncovering deeper insights into the underlying physics governing particle interactions.

\section{Logistics of combining likelihoods}
\label{sec:logistics}

One of the central promises of EFT is that it provides a common language for describing physics beyond the standard model across different measurements and experiments.
This common language facilitates global statistical combinations, which exploit the capability of different analyses to constrain orthogonal directions in Wilson coefficient space.

Statistical combinations in high energy physics have a long history, especially with regards to precision electroweak fits, the discovery of single top quark production, and the discovery of the Higgs boson.
These combinations led to tremendous technical development of combination tools and procedures, but EFT combinations present some unique challenges, which will require additional development.
It is crucial that these tools be specified in a human-readable fashion, so that they can be used far into the future even though programming languages and execution environments will change.
It is also very important that the digital artifacts that become inputs to a combination have enough flexibility to permit changes in the underlying theoretical framework, for example the introduction of higher-dimension EFT operators or more precise calculation tools.

The time is ripe for this technical development, which will enable us to establish not only useful combination tools and procedures for EFT, but also to define the legacy of the HL-LHC and of future Higgs factories.

\subsection{Grand combinations at LEP and Tevatron}
\speaker{Tom Junk}
Searches for new particles and interactions grew in complexity
in the 1990s at LEP, with the use of neural networks and other
multivariate analysis techniques.  Statistical methods and software
kept pace, being able to use distributions of discriminant variables
for each event optimally, rather than just cutting and counting.
Multivariate analysis techniques however are typically optimized to
find a specific signal with an a priori model of the background.
Searching for more general signals requires care to make sure the
efficiencies stay high and are well understood, and that separation
from backgrounds remains close to optimal.

The combinations of multivariate Higgs boson search results within each of the
four LEP collaborations, ALEPH, DELPHI, L3, and OPAL, were necessary
in order to produce publishable results.  Combinations of the four
collaborations' results provided the most sensitivity to the sought-after
signals.  Preparing digital artifacts for inputs to the combination
was organized by a working group that standardized the formats and
set deadlines.  The format was designed so that choices that
had an impact on the combined results, except the choice of the
final statistical method, were made by the experimental collaborations
and not by the combiners.  Issues of binning and interpolation methods
were handled by exchanging Fortran code which could read in external
data files as needed.  Systematic uncertainties were provided by named
source, where sources with the same name were considered 100\% correlated
and sources with different names were considered uncorrelated.
Arbitrarily-correlated uncertainties could be decomposed into
correlated and uncorrelated pieces.

Combinations at the Tevatron increased in complexity, mainly due to the
need to handle larger systematic uncertainties and the shift to
ROOT and C++.  Single Top and Higgs combinations followed similar
patterns to those used at LEP, with increased attention to shape
uncertainties and profiling the likelihood in the calculation of the
test statistic.  As was the case with the LEP experiments, each
collaboration provided inputs, but also each collaboration performed
all of the combinations, reproducing their own and those of the other
collaboration(s).  Results could only be approved if they were in
agreement among several combiners.  The necessary checking of the
inputs and reproduction of the results was a valuable tool for finding
and correcting mistakes in the inputs or the interpretations of them.
Better tools for preparing, exchanging, combining, and preserving
results have been created for use at the LHC, expanding on experience
gained at LEP and the Tevatron.

\subsection{Grand combinations at HL-LHC and FAIROS-HEP}
\label{sec:cranmer}
\speaker{Kyle Cranmer}
The gold standard for EFT results should be based on statistical combinations similar to those used for the Higgs discovery, i.e., statistical models of the observations without unfolding. The infrastructure for these combinations is mature, but we lack an agreed upon standard for describing how the distribution of a specific process (e.g. `signal') depends on the EFT coefficients. Recent experience has shown that there are significant advantages to separating the mathematical specification from the implementation, and that this facilitates efforts to publish these statistical models. Such a specification could be a straight-forward extension of the specifications used for binned-template fits such as the HistFactory / pyhf specification or what is used in the CMS Combine tool. It would be natural to have a specification for both the linear and quadratic EFT expansion strategies.  Finally, in order to extend the results to new EFT operators, update background modeling, or incorporate the effect of EFT operators on backgrounds, a RECAST-like service for EFTs would be valuable. In the case of new EFT operators, this can be achieved very efficiently through reweighting, and the service can export a new statistical model.

\subsection{SMEFT for top quarks at LHC and future Higgs factories}
\speaker{Michael Peskin}
One of the best opportunities for the discovery of
Beyond-Standard-Model effects at colliders that we know how to build
comes in the precision study of the top quark.  This motivates a
detailed search for nonzero SMEFT coefficients for one or more of the
dimension 6 SMEFT operators associated with the top quark.

Different BSM models
treat the top quark very differently.   In some models, such as
supersymmetry and two-Higgs-doublet models, the top quark is an
ordinary quark with an order-1 Yukawa coupling.   Its form factors are
affected by electroweak perturbative corrections, of typical size
$(\alpha_w/4\pi)(v^2/M^2)$, where $\alpha_w$ is the weak interaction
$SU(2)$ coupling, $v = 246$~GeV is the Higgs boson scale, and $M$ is
the mass scale of BSM physics.   Other models --- in particular,
models in which the Higgs boson is composite --- require a stronger
interaction with the top quark in order to produce the large top quark
mass.  In these models, the BSM corrections are order-1 times
$v^2/M^2$, and often with a large numerical coefficient. In such
models, the top quark is said to be ``partially composite''.  The
phenomenology of the top quark interactions with a composite Higgs
sector presents real opportunities for discovery~\cite{Franceschini:2023nlp}.  These
ought to be taken more seriously  by LHC
experimenters.

Even considering only dimension 6 operators, SMEFT introduces a large
number of possible Lagrangian interactions for the top quark.   These
divide into three classes:   4-fermion operators solely within the 3rd
generation, 4-fermion operators that link the 3rd generation with
lighter generations, and operators that couple 3rd-generation
fermion bilinears to bosonic operators.   In the standard analysis for
the LHC~\cite{Hartland:2019bjb}, the number of these operators is 11 + 14 +
9 = 34.  The multiplicity of 4-fermion operators comes from the fact
that individual operators are current-current interactions with
specific helicities in the initial and final states, e.g.   $(\bar q_L
\gamma^\mu q_L)(\bar t_R \gamma_\mu t_R)$, and possibly with weak
isospin or $SU(3)$ color currents.   At an $e^+e^-$ Higgs factory
operating above the top quark threshold, the number of operators is
also large~\cite{Durieux:2022cvf}, seven 4-fermion operators coupling the
3rd generation quarks to electrons, and the above 9 fermion-boson
operators.  It will clearly be important to make joint fits to these
two data sets.  It should be noted that the 4-fermion contact
interactions involving first-generation quarks and those involving the electron  are
distinct but should be  similar in size. A global fit might include an
explicit  theoretical model to relate these interactions.

The most important effects of composite Higgs physics should be from
vector resonances coupling to $t\bar t$ and from mixing of the top
quark with vectorlike top quark partners.   The first class of effects
should show up in 4-fermion contact interactions, in resonant
enhancement of the couplings of $t$ to $Z$ and $W$  (associated with
SMEFT
operators
such as $(\Phi^\dagger \Dlr_\mu \Phi)(\bar t_{L,R}\gamma^\mu t_{L,R})$),  and
in 
resonant enhancement of the top quark Yukawa coupling (associated with
the SMEFT
operator $|\Phi|^2 \bar t_L\tilde \Phi t_R$).  The second effect is
visible in decrements of the top quark couplings to $Z$ and $W$ and in
loop effects decreasing the $Hgg$ coupling and the overall scale of
Higgs boson couplings.

The HL-LHC will offer a very large data set to explore for these
effects, with approximately 3 billion top quark pairs expected.  For
precision studies, the modeling of these events to extract angular
distributions and top quark spins will be challenging.

\begin{figure}
\begin{center}
\includegraphics[width=0.70\hsize]{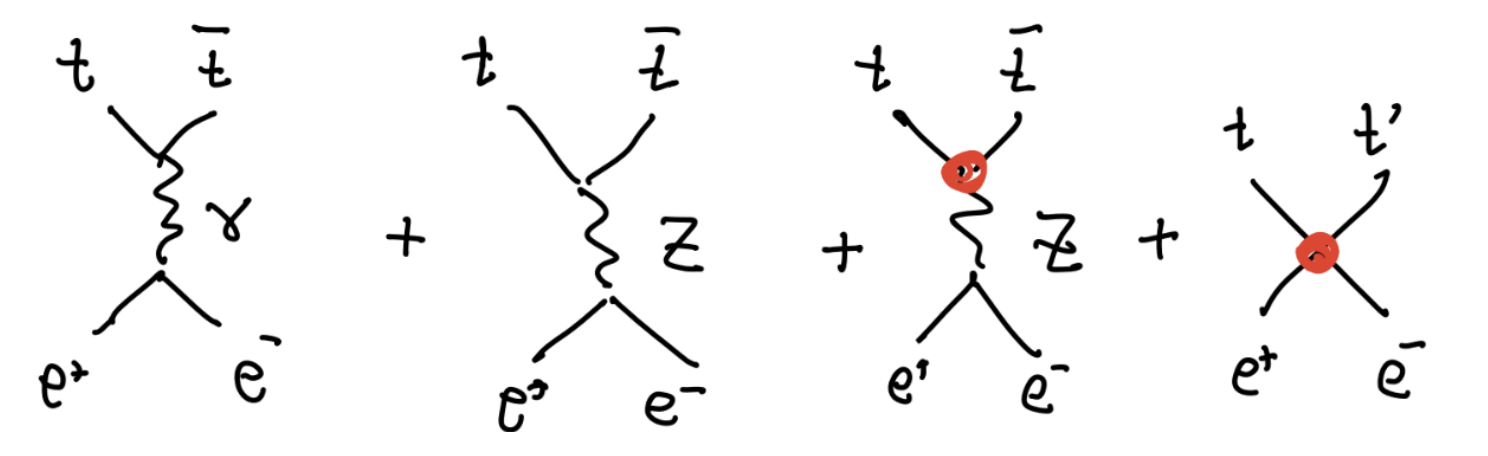}
\end{center}
\caption{Diagrams contributing to the cross section for $e^+e^-\to t\bar
  t$ at the leading order.}
\label{fig:eett}
\end{figure}

$e^+e^-$
colliders offer lower event samples but in a setting that makes it
easier to distinguish the effects of the various SMEFT operators.  In
particular, a linear collider can have high electron beam
polarization.  This makes it possible to measure 6 independent cross
section observables -- the forward, central, and backward cross
sections for each of two polarization settings. These are expected to
be measured at the parts-per-mil level of accuracy.  Adding positron
polarization as in the ILC design enhances this separation and
provides an additional tool for background
reduction~\cite{Amjad:2015mma}.  An $e^+e^-$ collider is
  particularly sensitive to the $t\bar t Z$ couplings, since this is
  part of the top quark pair production mechanism;  see Fig. 1.
One difficulty is that, at a fixed energy, the effect of the two
$t\bar t Z$ operators are degenerate with the effects of 4-fermion
operators. Disentangling these operators requires running at two
different center of mass energies above the top quark threshold (e.g.,
550~GeV and 800~GeV).
  
I would like to add one simple comment on the archiving of LHC SMEFT
fits.  I am one of those conservative people who does not believe in
fits with dimension 6 operators including both linear and quadratic
terms.  The quadratic terms, being positive while dimension 8
operators can give negative terms, lead to constraints on Wilson
coefficients that have no physical basis.  A better strategy is to
fit with linear terms in the dimension 6 Wilson coefficients only and
construct the inverse covariance matrix $(C^{-1})_{ab}$, where $a,b$
run over the full set of Warsaw basis operators.    This matrix has
many zeros and, typically, is not invertible.  Truncating this matrix
to its positive subspace and then inverting gives the constraints on
those operators that the data actually constrains.   When additional
data sets are added, the
matrix $C^{-1}$ will be invertible over a larger subspace, and so more
directions will be constrained.  It will be useful to quote two errors
on $C^{-1}$, a direct experimental error and an error due to
truncation of $1/M^4$ terms (as proposed in Sec.~\ref{dim8_as_uncertainty}). 

\subsection{Discussion}

The history of grand combinations in collider physics details:
\begin{itemize}
    \item A tightly coupled dialogue between theorists and experimentalists that defines the mathematical language in which we should summarize tests of wide classes of limits on new physics and corresponding measurements;
    \item The evolving complexity of computational techniques for statistical inference to address experimental uncertainties in requisite detail; and
    \item The accessibility and ease of interpretation of these analysis data products for theorists, experimentalists, and scientists in other sub-fields of particle physics.
\end{itemize}
In this session we discussed the past, present, and future of these aspects from the perspective of both physics, software, and even sociology.

Beginning with the practical matter of exchanging and combining physics results, all three presentations provided unique insight on what needs to be achieved in a multifaceted analysis product like a grand EFT fit combination.
The discussion of the history of combinations made a clear case about striving to maintain simplicity of components of the combination so that cross-checks are easy to perform and that the overall pieces of the combination each themselves make sense.
This goal was more easily attained in the era of Fortran based analyses of LEP where data interchange formats were necessarily simple and restricted, such as custom text files.
However, as technology evolved and resulted in the rise in popularity and utility of C++ during the Tevatron, this goal was challenged by the advent of bespoke file formats and increasing computing power since the number of viable methods to achieve the goal of a combination greatly increased. 
Accumulating experimental information became necessarily more complex and started to require standardization, which has persisted to the LHC experiments, and will be required for combinations with the output of future colliders.
Special attention should be paid to making sure that analysis products, like the complete likelihoods and software to evaluate them, can be calculated as-is in future computing systems for combinations.

These grand combinations of the LHC data, much like the electroweak combinations of LEP, will stand as definitive results for multiple decades due to the time it takes to construct next-generation colliders and experiments, and then take sufficient data from them.
As we are able to constrain more parameters with new data and the field approaches being able to constrain the full parameter space, the directions to pursue and subsequent machines to build will become better justified.
Useful and guiding summaries like the inverse covariance matrix for all EFT parameters, and the ability to reproduce and combine them on demand to update our knowledge, will be essential tools in determining the course of the collider physics in the next century.
While the inverse covariance matrix is certainly the most robust data product to changing software environments and computing techniques, it is the result that is least rich in experimental information and hardest to precisely combine with new measurements as information about systematic uncertainties and correlations is lost.
This means that we need to invest in analysis preservation infrastructure for EFT results, and that it will be critical to the use of LHC data products throughout the future of the field.

\section{Unfolded results}
\label{sec:unfolded}

Measured distributions in particle physics are distorted by the finite resolution
and limited acceptance of the detectors.
The transformation to the underlying
true distribution, called unfolding, is crucial for estimating the EFT sensitivity of certain variables. Limitations can arise due to our unfolding capabilities and methods, which should be considered in the EFT interpretation.
This session discusses the limitations
and the unfolding methods used in physics analyses. In addition, the treatment of the fiducial measurements that can be used for EFT interpretations is highlighted.
Both theoretical and experimental presentations were included and gave the current picture of unfolding measurements and treatment, emphasizing in the lessons learned so far and opening the discussion for a future Run3-combination.
The main aspects of the topic were presented. Section~\ref{chal_EFT_andr}
focus on the challenges in the EFT interpretation of the unfolded cross-section measurements, followed by section~\ref{sec:langford} on the Higgs and Simplified Template Cross Sections. Finally, on Section~\ref{sec:petriel} we focus on SMEFT probes with LHC Drell-Yan data.

\subsection{Challenges in the EFT interpretation of unfolded cross section measurements}
\label{chal_EFT_andr}
\speaker{Andrew Gilbert}
Unfolded differential cross-section measurements have traditionally been a crucial interface between theory and experiment, and can readily be used for reinterpretation with EFT. A key advantage is that such measurements do not require any particular new physics model at the time they are produced. As long as the definition of the fiducial selection and unfolded observables are preserved (e.g., via a RIVET plugin), a new interpretation can be produced potentially decades later. There are, however, some challenges. We are typically only able to unfold a small number of variables simultaneously, meaning we may not have optimal sensitivity to all operators. Backgrounds which may be EFT sensitive are assumed to follow standard model distributions and subtracted. We are often interested in the high energy tails of distributions, which are difficult to unfold when the expected number of events is small. Furthermore, there is an implicit assumption that the EFT effects do not modify the efficiency times acceptance within each bin. With the LHC experiments in the process of releasing the full statistical model information (from which the unfolded cross sections are derived), we have a more powerful tool to address some of these issues. Nonetheless, releasing differential cross section results remains important for comparisons to theory, and offers a fallback method for future reinterpretation.

\subsection{Higgs and Simplified Template Cross Sections}
\label{sec:langford}
\speaker{Jonathon Langford}
The LHC Run 2 dataset has enabled the CMS and ATLAS experiments to go beyond inclusive measurements of Higgs boson production and decay, and begin measuring Higgs boson interactions differentially.  One approach is within the Simplified Template Cross Section (STXS) framework, where Higgs boson events are first split by production mode and secondly by kinematic variables such as the transverse momentum of the Higgs boson or the number of additional jets. Both experiments have performed STXS measurements in a number of the major Higgs boson decay channels, thus developing a granular, kinematic description of Higgs boson production. One of the key advantages of STXS is that it provides a natural framework for BSM interpretations, including the use of Standard Model Effective Field Theory (SMEFT). The standard approach here is to parameterize the Higgs boson cross sections, at the granularity of the STXS, and decay rates as functions of the SMEFT Wilson coefficients, which enter the likelihood as signal-strength modifiers. In the talk, the caveats of this approach were discussed which arise from the fact that we cannot fully encapsulate all EFT effects within simple rate scaling functions. These caveats include acceptance corrections due to no fiducial selection on the Higgs boson decay products in the STXS, selection effects where EFT effects can vary significantly within a single STXS bin, and shape effects where the EFT can change the shape of the fitted observable used for signal extraction (e.g. a multivariate classifier score). In addition, the STXS binning choice is not optimized for SMEFT sensitivity. Following this, possible future improvements to the STXS were discussed, including adding a fiducial selection on decay products, updating the binning scheme with finer granularity and improving the tools used to derive the parametrization (such as standalone reweighting after the detector simulation). Finally the use of STXS measurements within global EFT fits was presented along with a few points to be addressed. These include the choice of flavor scheme, the simultaneous parametrization of signal and background, and methods to ensure orthogonality between analyses.

\subsection{SMEFT probes with LHC Drell-Yan data}
\label{sec:petriel}
\speaker{Frank Petriello}
The experimental precision for a multitude of Drell-Yan observables is approaching the percent level. Future studies can take advantage of this high-precision data to search for subtle deviations from Standard Model (SM) predictions. A framework for heavy new physics searches in the absence of new particles is the SM effective field theory (SMEFT), which contains all operators consistent with SM symmetries and which assumes a mass gap to any new physics. Detailed studies of the existing Drell-Yan data and of simulated future HL-LHC data reveal a rich program of discovery in this channel. Observables such as the invariant mass and forward-backward asymmetry are measured precisely enough to reveal higher-order dimension-8 corrections in the SMEFT. Access to these terms can potentially discriminate between ultraviolet completions of the SMEFT even without direct observation of a new particle. This direction will take on a new dimension with high-precision transverse momentum distribution measurements at an HL-LHC. Angular distributions in the Drell-Yan process also provide a ``smoking gun" signature of dimension-8 effects that are ripe for investigation. Future experimental analyses should take advantage of these vast possibilities in the Drell-Yan process.

\subsection{Discussion}

The approach of presenting unfolded results for later interpretation has several drawbacks. The primary ones pointed out by the speakers in this session include the necessity of using only few observables (which cannot be sensitive to all SMEFT operators); the built-in assumptions regarding background processes (usually that they are SM); the impact of SMEFT on efficiency and acceptance (often neglected); and the challenge of defining bins such that SMEFT effects do not vary significantly within a single bin. Concerns were raised in the question session about producing unfolded results in multiple correlated observables. It was suggested that, since the same issue is faced in MC generator tuning, it could be possible to use Monte Carlo to understand these correlations. It was also emphasized that unfolded results must be produced in bins where the SMEFT expansion parameter remains small - this is something that should be made very explicit in all interpretations of unfolded results. 

Considering these substantial drawbacks, the question was then raised about the utility of unfolded results for re-interpretation by theorists. As demonstrated in the final talk of the session (by Frank Petriello), unfolded cross sections are very useful for theorists to perform interpretations without diving deeply into the details of ATLAS and CMS. The consensus was that unfolded results should continue to be produced, but full (reco-level) likelihoods should also be published and preserved. One alternative to re-interpreting unfolded results would be for the experiments to provide recipes for forward-folding. This would reach a similar result, but would require more work and experimental expertise from the person performing the interpretation. 

\section{Building a sample library}
\label{sec:library}
In planning the session, a few discussion items that are particularly relevant to event sample production were identified: the choice of UFO\footnote{Universal FeynRules Output} model and what diagram classes to consider (e.g.~order in $\alpha_s$); what theory systematic uncertainties are to be considered; the choice of starting point in model space to use for the initial event sample that may subsequently be re-weighted; and prospects for sharing samples across experiments. Speakers were invited to address any of the above points (or others), and a discussion followed.


\subsection{The interplay between PDF fits and heavy New Physics searches}
\label{sec:mantani}
\speaker{Luca Mantani}
The extraction of parton distribution functions (PDF) from data can potentially conceal effects of heavy new physics (NP), since the PDF parameterisation can be flexible enough to mimic the NP induced deviations in the tails of distributions. Moreover, expanding the kinematic coverage of the data at low-Q, for example by incorporating projected data from forward facilities, can help mitigate the NP ``contamination'' of the PDFs. This is because discrepancies between data at high-Q and low-Q would become apparent, prompting the exclusion of high-Q data from the global dataset. Ultimately, the capabilities of SIMUnet, a public framework designed for simultaneous global fits of PDFs and EFT Wilson Coefficients were presented. Through SMEFT PDF extraction, SIMUnet has the potential to disentangle NP effects in the data, even without the need for additional projected measurements.

\subsection{Adventures in producing EFT samples}
\label{sec:wightman}
\speaker{Andrew Wightman}
Monte Carlo reweighting is a powerful tool that enables exploring the high dimensional phase space of EFT parameters without needing to generate an unreasonable number of samples. A key part of this approach is the ability to extract quadratic parameterizations, which enables us to estimate event yields in any kinematic distribution and event selection for an arbitrary choice of EFT parameters.

There are already techniques and tools developed that implement much of the technical machinery needed to extract these quadratic parameterizations, such as WCFit, but these tools still need to be placed into CMSSW\footnote{CMS offline software framework} so that they can be more widely accessed within the collaboration. This will also allow for concrete decisions to be made about conventional choices regarding accessing and storing these objects throughout the different stages of producing a EFT oriented MC sample. It is also important to consider the choice of starting point used to generate the MC samples. Some preliminary checks of how well the sample reweights to different parts of the phase space should be done in order to get a proper estimate for how many events will be needed to ensure reasonable statistical power for an analysis.

\subsection{Building the STXS Parameterization}
\label{sec:knight}
\speaker{Charlotte Knight}
An EFT parameterization of the STXS is a big undertaking, requiring the use of multiple techniques and tools, e.g.~both reweighting and generation-based approaches. Furthermore, post-generation tools are necessary to properly account for acceptance effects, and to perform more validity checks (see Sec.~\ref{sec:langford}). To reduce the amount of duplicated work, a common parameterization is being developed in collaboration with CMS, ATLAS and theorists. This parameterization will be published in a proposed common format and it is our hope that future publications will conform to this so that a library of parameterizations can be created. We would also like to encourage, where possible, the release of workflows to accompany these publications so parameterizations can be easily reproduced and altered if desired.

\subsection{Discussion}

In discussing the interplay between PDFs and SMEFT (Sec.~\ref{sec:mantani}), there was a question about the adequacy of existing methods for estimating PDF uncertainties in event samples produced for use with EFT fits in CMS, and how to fold the results of SIMUnet studies into experimental uncertainty estimates. It was understood that there will be a need for improved PDF measurements using high-$x$ low-$Q^2$ data such as that provided by forward physics facilities (e.g.~FASER) that may better disentangle PDF degrees of freedom from EFT ones. SIMUnet could potentially be used to design a new PDF model for use in experimental fits that is more conservative in its extrapolations to high $Q^2$ in view of the sensitivity to NP effects in that region.
A potential action item for CMS analysts is to measure the degree of correlation between PDF Hessian eigenvariations and EFT parameter effects on their analysis regions and compare the result with that provided by SIMUnet for the same event topology, to better understand if the uncertainty estimates can detect for possible absorption of NP discrepancies into the PDF model.

During the discussion on sample generation (Sec.~\ref{sec:wightman}), several questions revolved around what an optimal generation strategy might look like. For example, it is unclear whether generating inclusive samples (e.g.~$t\bar{t}$) or exclusive decay mode samples with EFT weights is more resource-efficient. In the case of this analysis, samples were produced with exclusive decays generally. Should one use the same point in WC space for sampling events across different processes? Generally, no because it can be challenging to find a point in WC space that samples events that can be effectively reweighted to all interesting WC values. In the case of the analysis presented in \ref{sec:wightman}, one sample was sufficient but in other cases there may be a need to combine several samples together that are generated from different starting points. Methods for combining samples can be constructed, but a thorough evaluation of the options should be an action item for CMS analysts.
There is progress on integration of the code to compute the WC polynomial expansion into CMSSW, but it is not complete yet.

In the discussion on building the STXS parameterization (Sec.~\ref{sec:knight}), many in the audience were very interested in the post-generation reweighting technique presented. It was discussed how best to integrate the MadGraph-based reweighting module into vectorized python workflows and what common components are shared with the code for computing WC parameterization used in the previous talk that could be harmonized. The post-generation reweighting technique was identified as an effective method to determine the EFT parameterization of reducible backgrounds without having to re-generate potentially large simulation samples, provided it can be shown that the result is faithful to that of reweighting a dedicated sample. Even samples produced with a different generator could in principle be reweighted, but the results should be treated with caution, especially in the case where the considered diagrams are not common, and are likely not meaningful between LO and NLO QCD generators.
It was highlighted that, in cases where the quadratic term is to be used mainly as an estimate of the missing higher order terms in the $1/\Lambda$ expansion, that one can relax the requirements on the MC statistical uncertainty in determining those coefficients.
The need for a schema for the JSON data format for bin parameterizations was also highlighted.

\section{Conclusion}
As evidenced by the extensive discussions, a dedicated workshop could easily have been arranged for each of the five themes. Although some consensus on the best approach to performing large-scale experimental SMEFT combinations is building, it is clear that more work will be needed. We believe that there are paths laid out here that are actionable in the short term, while still being flexible to future changes in requirements for (re-)interpretability. Future workshops will help refine a plan of action for this community.

\section*{Acknowledgments}
We would like to thank the University of Notre Dame for making space available for this workshop. We would also like to thank the Fermilab LHC Physics Center for hosting the inaugural workshop and nucleating this community.

\bibliography{main.bib}
\bibliographystyle{unsrt}

\end{document}